\documentclass[10pt, letter, twocolumn]{IEEEtran}

\usepackage{graphicx}
\usepackage{epsfig}
\usepackage{algorithm}
\usepackage{algorithmic}
\usepackage{ifmtarg}
\usepackage[center]{caption}
\usepackage{footnote}
\usepackage{cite}
\usepackage{amssymb}
\usepackage{amsthm}
\usepackage[fleqn]{amsmath}
\usepackage{amsmath}
\usepackage{amsfonts}
\usepackage{epstopdf}
\usepackage{enumitem}
\usepackage{subfig}
\usepackage{color}
\usepackage{comment}
\usepackage{ulem}

\usepackage[utf8]{inputenc}
\usepackage[english]{babel}

\usepackage[font=normalsize]{caption}
\begin{document}
\title{Improvement of Bi-directional Communications using 
Solar Powered Reconfigurable Intelligent Surfaces}
\author{
\IEEEauthorblockN{Abdullah Almasoud$^{\text{1}}$, Mohamed Y. Selim$^{\text{2}}$, Ahmad Alsharoa$^{\text{3}}$, and Ahmed E. Kamal$^{\text{2}}$}\\
\IEEEauthorblockA{$^{\text{1}}$Department of Electrical Engineering, College of Engineering, Prince Sattam Bin Abdulaziz University, Al-Kharj 16273, Saudi Arabia, Email:am.almasoud@psau.edu.sa }\\ 
\IEEEauthorblockA{$^{\text{2}}$Iowa State University, Ames, Iowa, United States, Email: \{myoussef, kamal\}@iastate.edu}\\
\IEEEauthorblockA{$^{\text{3}}$Missouri University of Science and Technology, Email: aalsharoa@mst.edu} \vspace{0.5cm}\\
    {\textbf{(Invited Paper)}}

{\thanks {\vspace{-0.4cm}\hrule \vspace{0.1cm} \indent This research was supported in part by grant 1827211 from the National
Science Foundation, USA.}}
}
%\begin{center}
%\end{center}

\maketitle
\thispagestyle{empty}
\pagestyle{empty}

\begin{abstract}
\boldmath{Recently, there has been a flurry of research on the use of Reconfigurable Intelligent Surfaces (RIS) in wireless networks to create dynamic radio environments.
In this paper, we investigate the use of an RIS panel to improve bi-directional communications. Assuming that the RIS will be located on the facade of a building, we propose to connect it to a solar panel that harvests energy to be used to power the RIS panel's smart controller and reflecting elements. 
Therefore, we present a novel framework to optimally decide the transmit power of each user and the number of elements that will be used to reflect the signal of any two communicating pair in the system (user-user or base station-user). An optimization problem is formulated to jointly minimize a scalarized function of the energy of the communicating pair and the RIS panel and to find the optimal number of reflecting elements used by each user. Although the formulated problem is a mixed-integer non-linear problem, the optimal solution is found by linearizing the non-linear constraints. Besides, a more efficient close to the optimal solution is found using Bender decomposition.
Simulation results show that the proposed model is capable of delivering the minimum rate of each user even if line-of-sight communication is not achievable.
%Simulation results illustrate the performance of using RIS in improving the achievable data rate of the two-way relaying system.
}
\end{abstract}

\vspace{-0.1cm}
\section{Introduction}\label{Introduction}

Reconfigurable Intelligent Surfaces (RIS) is a reconfigurable meta-surface consisting of several passive reflecting elements and a smart controller. By modifying the amplitude and the phase of the incident radio waves, RIS can dynamically control the radio environment for various purposes, such as enhancing received signal and canceling interference \cite{EmilMyth}. The RIS will then play a role similar to the one played by a relay station using decode-and-forward or amplify-and-forward with beamforming and directional antennas, but at a lower cost, lighter weight, less power consumption, and no added interference.

RIS has many attractive advantages such as ease of deployment, spectral efficiency enhancement, and energy-efficiency. It only reflects the electromagnetic waves regardless of the technology used in transmitting these waves.
For these reasons, RIS constitutes a promising software-deﬁned architecture that could potentially enable telecommunication operators to sculpt the communication medium that comprises the network.

Several surveys covering RIS state of the art, principles, and opportunities have been published \cite{IRSsurvey1,IRSsurvey2,IRSsurvey3}.
There is a number of papers published in this area covering channel estimation, modeling, and measurements \cite{IRSchannel1,IRSchannel2,IRSchannel3}, and optimization and resource allocation \cite{IRSoptimization1,IRSoptimization2,IRSoptimization3,IRSoptimization4}. Also, comparing RIS with other technologies such as relays is addressed by \cite{IRSotherTech1,IRSotherTech2}.

Cooperative communications of relaying have been proposed in the literature to increase the overall aggregate data throughput, extend the network coverage area, and reduce the users' transmitted powers, hence, decreasing the interference to nearby users. Further, in the absence of the direct link, relays can maintain the communications links between non-line-of-sight users~\cite{HSlim}. %In contract to relaying, the RIS ...... 

While both relays and RISs serve conceptually similar purposes, the relay plays the role of receiving and retransmitting the signal with amplification. In \cite{RISRelay:wang}, the authors consider an RIS-assisted two-way
relay network in which two users exchange information via the
base station (BS) with the help of an RIS. They formulated an optimization problem where the minimum SNR of the two users is maximized under the transmit power constraint at the BS. The authors in \cite{RISRelay:ying} proposed an architecture consisting of two side-by-side RISs connected via a full-duplex relay. They proved that this architecture has the potential of achieving promising gains while requiring fewer reflecting elements. Finally, \cite{RISRelay:atapattu} investigated the two-way communication between two users assisted by an RIS. They considered two users communicating simultaneously over Rayleigh fading. They formulated an optimization problem to maximize the signal-to-interference-plus-noise ratio (SINR). They showed that their proposed greedy-iterative algorithm can achieve high performance in terms of spectral efficiency and low computational complexity. 

The key difference between a relay and an RIS is that a relay actively processes the incident signal before retransmitting an amplified version of the signal. At the same time,
an RIS reflects the incident signal without any amplification using passive beamforming. The relay achieves a higher signal-to-noise
ratio (SNR) at the cost of a pre-log penalty due to the two-hop
transmission. The authors in \cite{RelayRIS:huang} compared an amplify-and-forward relay with RIS showing that RIS is much more energy-efficient than the relay. Since decode-and-forward (DF) relay outperforms amplify-and-forward (AF) relay, the authors in \cite{RelayRIS:emil} compared a DF relay with RIS. The main observation in this paper is that very high rates and/or large number of reflecting elements are needed to outperform DF relay, both in terms of minimizing the total transmit power and maximizing the energy efficiency.

This paper considers the use of an RIS to improve communications between two wireless communicating devices that can be either two end users or a basestation and an end user.
The problem we consider is when the two communicating devices do not have a direct line-of-sight between them, then rather than using a relay station, which requires a line power connection, and also consumes a significant amount of power, an RIS panel may be used to provide this enhancement.
The RIS panel does not require a significant amount of power, and can be powered from a renewable energy source, e.g., using a solar panel, and can therefore be located at places where the relay stations may not be otherwise located.
The system model assumes that users are tuned to different channels, and different elements of the RIS panel are used to support communication in one direction only, but collectively they support communications in two directions.
The paper formulates an optimization problem to minimize the energy consumed by end stations, while taking into consideration the energy harvested by the solar panel used to power RIS panel's microcontroller(s) and elements.
The optimization problem determines the number of RIS panel elements necessary for operation, and which RIS panel elements are used for communication in each of the two directions, as well as the end users transmit powers.
A minimum transmission rate constraint is enforced.
The optimization problem is a mixed-integer non-linear problem.  
The non-linearity is removed by linearizing the non-linear constraints.
Moreover, a close to optimal heuristic solution based on Bender's decomposition is introduced.
Numerical results explain the performance of the system.

The paper is organized as follows.
In Section II we introduce some use cases that can be adapted to the of model of this paper.
Section III introduces the system model.
The optimization problem formulation is introduced in Section IV as well as the Bender decomposition solution approach.
Section V  introduces simulation results, and Section VI concludes this paper with some remarks.

\section{Use Cases}
In this section, we present use cases that integrate RIS in the data communications scheme.
% for different models or scenarios. The RIS is used to enhance the data rate between users.
Given that our proposed scheme introduces a general communication model between a source and a destination assisted by an RIS panel, this model can be applied to several use cases. 
We present some of them below.

\subsection{D2D-RIS Use Case}
Direct communications can improve the spectrum utilization and enhance the data throughput between nearby users~\cite{D2D_su,dynamic}. Several short-range wireless frameworks have been proposed in the literature to allow nearby users to communicate directly with each other such as Bluetooth, WiFi-Direct, LTE-Direct, and UAV-Direct~\cite{D2D_su,Alsharoa_C20}. The major differences between these frameworks are the communication ranges, applications, and discovery mechanisms. Although these techniques have larger discovery ranges than self-discovery (without the help of base station), it is sometimes challenging in infrastructure-less environments (no or minimum access to the base station). The challenge is how to manage the resources between the D2D pairs in an efficient way.
%\textbf{WHAT IS CHALLENGING EXACTLY?}

The RIS using LTE-direct can enhance the D2D link~\cite{R12}. The D2D enhancement model can be given as:
\begin{itemize}
 \item D2D users communicate with each other via a direct link and an RIS link. One of the users can be a mobile device, while the other can be another mobile device or a base station.
 \item The user's discovery is done using the LTE direct technique. 
 \item The base station manages (without being involved in the data communications) the bandwidth resources and transmit power allocations to reduce the interference between D2D users, and therefore increases the aggregate data rate.
 \item RIS optimizes the cells/elements assignment between different D2D pairs using its controller.
\end{itemize}

\subsection{BS-RIS-BS Use Case}
In this use case, one BS is providing the fronthaul/backhaul connectivity to another BS. If the direct link between the two BSs suffers from shadowing or even failures, the RIS will play an important role in providing an alternate path for the fronthaul/backhaul connectivity between the two BSs. The authors in \cite{RISBackhaul:aljarrah} considered the deployment of RIS for wireless multi-hop backhauling of multiple BSs. The performance of the proposed scheme is evaluated in terms of outage probability. They showed that the proposed scheme has several desired features that can be exploited to overcome some of the backhauling challenges, such as severe attenuation.

\subsection{BS-UAV$^{RIS}$-USER Use Case}
In this use case, the RIS is mounted on a UAV, i.e., UAV$^{RIS}$, to assist the communication between two communicating devices, i.e., BS-USER, BS-BS, USER-USER. This configuration can improve the channel quality in urban environments given the high probability of line-of-sight that the UAV can achieve. The authors in \cite{RISUAV:Lu} proposed a new 3D networking scheme enabled by UAV$^{RIS}$ to achieve panoramic signal reflection from the sky. They showed that UAV$^{RIS}$ not only enjoys higher deployment flexibility but also is able to achieve 360 degrees panoramic full-angle reflection.

%show that the proposed design can achieve significant performance gain

\subsection{UAV-RIS-BS/USER Use Case}
In this use case, the RIS is mounted on a building's facade and assisting the communication between the UAV and the BS or the UAV and the user. The authors in \cite{RISUAV:li} demonstrated that the proposed configuration can considerably improve the system's average achievable rate.

\section{System Model}\label{SystemModel}
In this section, we investigate a time-slotted system of a finite time period divided into $\textbf{T} = 1,..,T$, time slots of equal duration $T_s$.

\subsection{Network Model}
\begin{figure}[h!]
  \centerline{\includegraphics[width=3in]{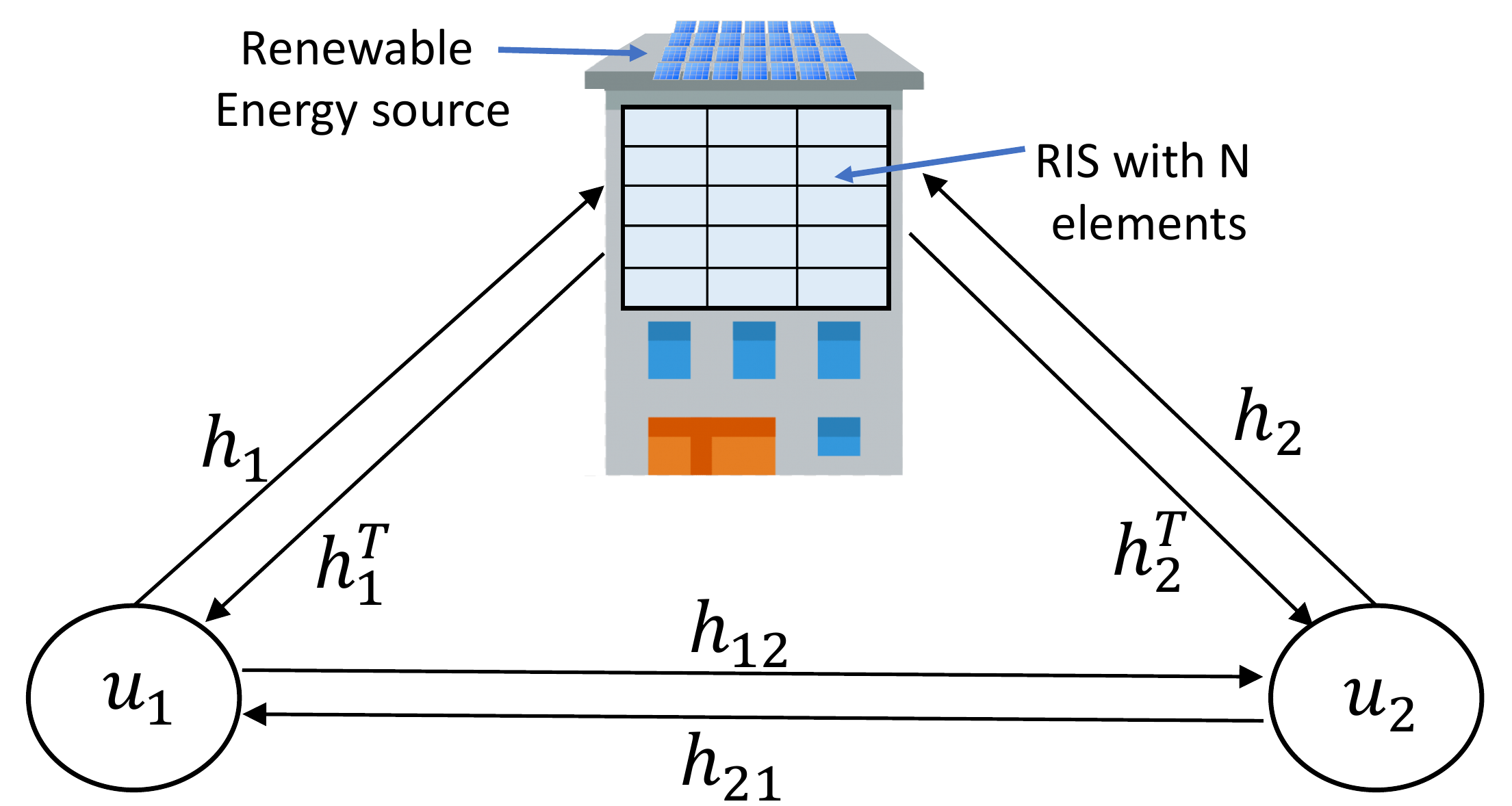}}
   \caption{\, System model of two-way relaying.}\label{fig1}
	\vspace{-0.3cm}
\end{figure}

We considers a two ways communications system where two battery-operated single antenna users, denoted by $u_1$ and $u_2$, exchange information through the help of the RIS containing $N$ reflecting elements as shown in Fig.~\ref{fig1}. We assume that $u_1$ and $u_2$ transmit their signals using different bandwidth sets, namely $B_1$ and $B_2$, respectively.
Also, we assume that the RIS panel is equipped with reflecting elements responsible of reflecting the received signal from u$_1$ to u$_2$ and vise versa.% with $B_2$ and from u$_1$ to u$_2$ with $B_2$ with $B_2$.

We assume NLoS transmission based on the reflection from the RIS panel in addition to the LoS transmission. Both $u_1$ and $u_2$ transmit their messages $x_1$ and $x_2$ simultaneously at time $t$ to all RIS elements with a power denoted by $P_1(t)$ and $P_2(t)$, over $B_1$ and $B_2$, respectively. Therefore, the consumed transmitted energy at $u_1$ and $u_2$ during time $t$ are given, respectively, as $E^U_1=T_s P_1(t)$ and $E^U_2=T_s P_2(t)$.
Each RIS element can take one of the following action: a) towards $u_1$: reflect the received message from $u_2$ to $u_1$ over $B_2$, b) towards $u_2$: reflect the received message from $u_1$ to $u_2$ with $B_1$, or c) no-action: no reflection and the element will be turned off to reduce the energy consumption. In other words, due to energy constraint, some of the RIS elements will help in reflecting the received signals and the other elements will be turned off.
Let us define $P^\text{max}$ as the peak power at $u_1$ and $u_2$, respectively.
We defined the LoS channel gain between $u_1$ and $u_2$ as $h_{12}$.
The channel gain between $u_1$ and RIS, and the channel gain between $u_2$ and RIS are defined, respectively, as $h_1$ and $h_2$ 
While, the reverse channel gain between RIS with  $u_1$, and $u_2$ are denoted by $h^T_1(t)$ and $h^T_2(t)$, respectively, where $(^T)$ denotes the hermitian operator.
Without loss of generality, all channel gains are assumed to be constant during the two transmission phases.
Also, all noise variances are assumed to be equal to $\sigma^2$, and $\mathbb{E} \left[|x_{1}|^{2}\right]=\mathbb{E} \left[|x_{2}|^{2}\right]=1$, where $\mathbb{E}\left[\cdot\right]$ denotes the expectation operator.

\subsection{Data Rate}

The LoS received signals at $u_1$ and $u_2$ are given, respectively, as:
\begin{equation}
    y^{LoS}_1(t)=h_{21}\sqrt{p_2(t)} + n_1,
\end{equation}
\begin{equation}
    y^{LoS}_2(t)=h_{12}\sqrt{p_1(t)} + n_2,
\end{equation}
where $n_i$ is the noise at $u_i$. 
Let us now define a binary variable $\epsilon_{i,n}$ equal to 1 if the element $n$ reflects the signal towards $u_i$ during time slot $t$ and 0 otherwise and given as:
\begin{equation}\label{ep_eq}
   \epsilon_{i,n}(t)=
   \begin{cases}
   1, & \text{if the element $n$ reflects the signal towards $u_i$,}\\
   0, & \text{otherwise.}
    \end{cases}
\end{equation}
Note that, each RIS element will be used either to reflect to the fRISt user of the second one. Therefore, the following condition must be satisfied:
\begin{equation}
    \sum\limits_{i=1}^2 \epsilon_{i,n}(t) \leq 1, \quad \forall n, \forall t,
\end{equation}

The following diagonal matrix represent the properties of the RIS corresponding to $u_i$:
\begin{equation}
    \mathbf{\Phi}_i(t) =  \, \alpha ~diag(\epsilon_{i,1}(t)e^{j\phi_1}(t), \dots , \epsilon_{i,N}(t) e^{j\phi_N}(t))
\end{equation}
where $\alpha \in (0,1]$ denoted as the amplitude reflection coefficient and $\phi_n(t)$ is the phase shift variables of element $n$ that is controlled and optimized by the RIS.
Therefore, the received reflected signal to $u_1$ and $u_2$ are given, respectively, as:
\begin{equation}
     y^{RIS}_1(t) = (\mathbf{h}_1^T(t) \mathbf{\Phi}(t) \mathbf{h}_2(t))\sqrt{p}_2(t) + r_1(t),
\end{equation}
\begin{equation}
    y^{RIS}_2(t) = (\mathbf{h}_2^T(t) \mathbf{\Phi}(t) \mathbf{h}_1(t))\sqrt{p}_1(t) + r_2(t),
\end{equation}

The data rate at $u_1$ and $u_2$ are given as:
\begin{align}
    &R_i(t) = \underset{\Phi}{\text{max}} \log_2 \bigg(1+ \frac{p_{\hat{i}}(t) |h_{{\hat{i}}i}+\mathbf{h}_i^T(t) \mathbf{\Phi_i}(t) \mathbf{h}_{{\hat{i}}}(t)|^2}{\sigma^2}\bigg) \label{R0}\\ 
\hspace{-1cm}  = & \log_2 \bigg(1+\frac{p_{\hat{i}}(t)(|h_{{\hat{i}}i}|+\alpha \sum_{n=1}^N \epsilon_{i,n}(t) |\mathbf{h}_{i,n}(t) \mathbf{h}_{{\hat{i}},n}(t)|)^2}{\sigma^2}\bigg)\label{R1}
\end{align}

where ${\hat{i}}=1$ if $i=2$ and vice versa.
Note that (\ref{R1}) can be derived from (\ref{R0}) based on the proof in \cite{Wu2}. It is worth pointing out that $\mathbf{h}_i^T(t) \mathbf{\Phi_i}(t) \mathbf{h}_{{\hat{i}}}(t) = \alpha \sum_{n=1}^N \epsilon_{i,n} |\mathbf{h}_{i,n}(t) \mathbf{h}_{{\hat{i}},n}(t)|$.

\begin{comment}
\small
\begin{align}
    &R_1(t) = \underset{\Phi}{\text{max}} \log_2 \bigg(1+ \frac{p_2(t) |h_{21}+\mathbf{h}_1^T(t) \mathbf{\Phi_1}(t) \mathbf{h}_2(t)|^2}{\sigma^2}\bigg) \label{R0}\\ 
    = & \log_2 \bigg(1+\frac{p_2(t)(|h_{21}|+\alpha \sum_{n=1}^N \epsilon_{1,n}(t) |\mathbf{h}_{1,n}(t) \mathbf{h}_{2,n}(t)|)^2}{\sigma^2}\bigg)\label{R1}
\end{align}

\small
\begin{align}
    &R_2(t) = \underset{\Phi}{\text{max}} \log_2 \bigg(1+ \frac{p_1(t) ||h_{12}|+\mathbf{h}_2^T(t) \mathbf{\Phi_2}(t) \mathbf{h}_1(t)|^2}{\sigma^2}\bigg)\\ 
    = & \log_2 \bigg(1+\frac{p_1(t)(|h_{12}|+\alpha \sum_{n=1}^N \epsilon_{2,n}(t) |\mathbf{h}_{2,n}^T(t) \mathbf{h}_{1,n}(t)|)^2}{\sigma^2}\bigg)\label{R2}
\end{align}
\normalsize
\end{comment}

\subsection{Energy Model}
In this paper, we assume that the RIS can harvest from Renewable Energy (RE) such as solar. We model the RE stochastic energy arrival rate as a random variable $\Theta$ Watt defined by a probability density function (pdf) $f_\Theta(\theta)$. For example, for photovoltaic energy, $\Theta$ can be interpreted as the received amount of energy per time unit with respect to the received luminous intensity in a particular direction per unit solid angle.

%The RIS energy consumption during time slot $t$ can be given by the following:

The power consumption of the RIS panel relates to the number of reflecting elements, $N$, and the bit resolution of the phase control. Since all the elements of the RIS panel are theoretically passive, they still consume energy in practical when the diodes used in each reflecting elements are ON \cite{RISpower}. However, this power is negligible compared with the power consumed by the smart controller. This power depends on the controller circuit implementation and the communication module used. Hence, it is reasonable to assume that the bit resolution of all phase shifters are identical on one RIS panel. Therefore the power consumption of the RIS is given by $NP_e$, where $P_e$ denotes the  dissipated  power  per  RIS  element  caused  by the circuitry required for the adaptive phase shift. Given that some elements may not be used, the RIS energy consumption during time slot $t$ can be given by the following:
\begin{equation}\label{consumed}
E^\text{RIS}_C(t)=T_s \sum\limits_{i=1}^2\sum\limits_{n=1}^N \epsilon_{i,n}(t) P_e 
\end{equation}
where $T_s$ is the time slot slot duration. The harvested energy at the end of time slot $t$ can be given as
\begin{equation}
E_H(t)=T_s \eta \, \theta(t),
\end{equation}
where $\eta$ is the energy conversion efficiency coefficient, where
$0\leq~\eta~\leq 1$.
Notice that the current stored energy in the RIS depends on the current harvested energy during slot time $t$, the previously stored energy during previous slots, and the energy consumption during time slot $t$.
Therefore, the stored energy in RIS at the end of time $t$ based on harvest-store-use model is given by
\begin{equation}\label{stored}
E_S(t)= \left[E_S(t-1)+E_H(t)- E^\text{RIS}_C(t)\right]^+, \forall l, \forall t,
\end{equation}
where $[x]^+=\max(0,x)$. 
The following constraint need to be respected to ensure that the harvested energy cannot exceed the battery capacity as
\begin{equation}\label{stored2}
E_S(t-1)+E_H(t) \leq \bar{E}_S, \forall t,
\end{equation}
In addition, to ensure the causality energy constraint (i.e., the RIS cannot consume energy more than the available energy in its battery at time $t$), we impose the following condition:
\begin{equation}\label{stored2}
E^\text{RIS}_C(t) \leq E_S(t-1), \forall t,
\end{equation}

The energy consumption at user $u_i$ for each $t$ is given as:

\begin{equation}
    E^U_i(t) = T_s p_i(t)
\end{equation}

\vspace{-.2cm}
\section{Problem Formulation}\label{problem}

\begin{equation}\label{R maximization2}
    \underset{\epsilon_{i,n}(t),p_i(t) \geq 1}{\text{minimize}} \quad \zeta E^\text{RIS}_C(t) + (1-\zeta) \sum\limits_{i=1}^2 E^U_i(t)
\end{equation}
subject to:
 \begin{equation}\label{peakP}
 p_i(t) \leq P^\text{max}, \quad \forall t, \forall i=1,2.
 \end{equation}
  \begin{equation}\label{assoc}
\sum\limits_{i=1}^ 2 \epsilon_{i,n}(t) \leq 1, \quad \forall n, \forall t,
 \end{equation}
  \begin{equation}\label{rate}
R_i(t) \geq R_{th} \quad \forall t, \forall i=1,2.
 \end{equation}
  \begin{equation}\label{energy1}
E^\text{RIS}_C(t) \leq E_S(t-1), \forall t,
 \end{equation}
  \begin{equation}\label{energy2}
E_S(t-1)+E_H(t) \leq \bar{E}_S, \forall t,
 \end{equation}
where $\zeta$ is the weight coefficient.

\subsection{Solving The Optimizaton Problem Optimally}\label{optimalsolution}

For simplification, consider that $|\mathbf{h}_{i,n}(t) \mathbf{h}_{{\hat{i}},n}(t)| = \mathbf{H}_{i,{\hat{i}},n}$. Then \eqref{rate} can be written as:

\small
\begin{equation}\label{Rate_threshold}
    \log_2 \bigg(1+\frac{p_{\hat{i}}(t)(|h_{{\hat{i}}i}|+\alpha \sum_{n=1}^N \epsilon_{i,n}(t) \mathbf{H}_{i,{\hat{i}},n})^2}{\sigma^2}\bigg) \geq R_{th}
\end{equation}
\normalsize

\small
\begin{equation}
     \bigg({p_{\hat{i}}(t)(|h_{{\hat{i}}i}|+\alpha \sum_{n=1}^N \epsilon_{i,n}(t) \mathbf{H}_{i,{\hat{i}},n})^2}\bigg) \geq (2^{R_{th}}-1){\sigma^2}
\end{equation}
\normalsize

Expanding the terms of the LHS will result in the following:

\small
\begin{align}\label{rateexpand}
    &\Bigg(p_{\hat{i}}(t)\Big(|h_{{\hat{i}}i}|^2 + 2 \alpha |h_{{\hat{i}}i}| \sum_{n=1}^N \epsilon_{i,n}(t) \mathbf{H}_{i,{\hat{i}},n} \nonumber\\
    &+ \alpha^2 \big(\sum_{n=1}^N \epsilon_{i,n}(t) \mathbf{H}_{i,{\hat{i}},n}\big)^2 \Big)\Bigg) \geq (2^{R_{th}}-1)\sigma^2
\end{align}
\normalsize

The term $\big(\sum_{n=1}^N \epsilon_{i,n}(t) \mathbf{H}_{i,{\hat{i}},n}\big)^2$ can be further expanded as follows: 

\small
\begin{align}
    &\big(\sum_{n=1}^N \epsilon_{i,n}(t) \mathbf{H}_{i,{\hat{i}},n}\big)^2 = \sum_{n=1}^N \epsilon_{i,n}(t) \mathbf{H}_{i,{\hat{i}},n}^2 \nonumber\\
    &+ \sum_{x=1}^N \sum_{\substack{y=1\\ y \neq x}}^N \epsilon_{i,x}(t) \epsilon_{i,y}(t) \mathbf{H}_{i,{\hat{i}},x}\mathbf{H}_{i,{\hat{i}},y}
\end{align}
\normalsize

Then \eqref{rateexpand} can be written as:

\small
\begin{align}\label{rateexpandnonlinear}
    &p_{\hat{i}}(t)|h_{{\hat{i}}i}|^2 + 2 \alpha |h_{{\hat{i}}i}| \sum_{n=1}^N p_{\hat{i}}(t)\epsilon_{i,n}(t) \mathbf{H}_{i,{\hat{i}},n} + \nonumber\\
    & \alpha^2 \bigg( \sum_{n=1}^N \underbrace{p_{\hat{i}}(t)\epsilon_{i,n}(t)}_\text{non-linear1} \mathbf{H}_{i,{\hat{i}},n}^2 + \nonumber\\
    &\sum_{x=1}^N \sum_{\substack{y=1\\ y \neq x}}^N \underbrace{p_{\hat{i}}(t) \tilde{\epsilon}_{i,xy}(t)}_\text{non-linear2} \mathbf{H}_{i,{\hat{i}},x}\mathbf{H}_{i,{\hat{i}},y} \bigg) 
    \geq (2^{R_{th}}-1){\sigma^2}
\end{align}
\normalsize

where $\tilde{\epsilon}_{i,xy}(t)$ is a new binary variable used to linearize the product of the two binary variables $\epsilon_{i,x}(t)$ and $\epsilon_{i,y}(t)$. This linearization is well-know and can be represented by the following inequalities:

\small
\begin{equation}\label{lineartwobinary1}
    \tilde{\epsilon}_{i,xy}(t) \leq \epsilon_{i,x}(t)
\end{equation}
\begin{equation}\label{lineartwobinary2}
    \tilde{\epsilon}_{i,xy}(t) \leq \epsilon_{i,y}(t)
\end{equation}
\begin{equation}\label{lineartwobinary3}
    \tilde{\epsilon}_{i,xy}(t) \geq \epsilon_{i,x}(t) + \epsilon_{i,y}(t) - 1
\end{equation}
\normalsize

The frist two inequalities make sure that $\tilde{\epsilon}_{i,xy}(t)$ will equal to zero if one of the two binary variables $\epsilon_{i,x}(t)$ and $\epsilon_{i,y}(t)$ or both of them equal to zero. The three inequalities guarantee that $\tilde{\epsilon}_{i,xy}(t)$ will equal to one if and only if both $\epsilon_{i,x}(t)$ and $\epsilon_{i,y}(t)$ are equal to one.

The two non-linear terms shown in \eqref{rateexpandnonlinear} are having similar representation of one continuous variable, $p_{\hat{i}}(t)$, and one binary variable, either $\epsilon_{i,n}(t)$ or $\tilde{\epsilon}_{i,xy}(t)$. These two terms can be linearized by introducing a new decision variable for each of them.

We now will consider the general case of linearizing a continuous variable, $p_{\hat{i}}(t)$, multiplied by a a binary variable $\xi_{z,n}(t)$. In other words, we will linearize $p_{\hat{i}}(t)\xi_{z,n}(t)$. Hence, we introduce a new decision variable $\tilde{p}_{\hat{i},z,n}(t)$ to absorb the non-linearity resulting from multiplying $p_{\hat{i}}(t)$ and $\xi_{z,n}(t)$. This non-linearity can be linearized without any approximation by considering $\tilde{p}_{\hat{i},z,n}(t)=p_{\hat{i}}(t)\xi_{z,n}(t)$ where the following inequalities have to be respected:

\small
\begin{equation}\label{ineq1}
    {p}_i(t) \geq \Tilde{p}_{\hat{i},z,n}(t) \geq 0,
\end{equation}
\begin{equation}\label{ineq2}
    \Tilde{p}_{\hat{i},z,n}(t) \geq P^\text{max}\xi_{z,n}(t) - P^\text{max} + {p}_i(t)
\end{equation}
\begin{equation}\label{ineq3}
    \Tilde{p}_{\hat{i},z,n}(t) \leq P^\text{max}\xi_{z,n}(t)
\end{equation}
\normalsize

where $P^\text{max}$ is the maximum transmit power. The fRISt two inequalities ensure that $\Tilde{p}_{\hat{i},z,n}(t)$ value is between $\xi_{z,n}(t)$ and $p_{\hat{i}}(t)$. The third inequality guarantees that $\Tilde{p}_{\hat{i},z,n}(t)$ equals to zero if $\xi_{z,n}(t)$ equals to zero. Finally, the three inequalities guarantee that $\Tilde{p}_{\hat{i},z,n}(t)$ equals to $p_i(t)$ if $\xi_{z,n}(t)$ equals to one.

By considering this linearization and rewriting \eqref{rateexpandnonlinear}, the final linearized constraint can be given as:

\small
\begin{align}\label{rateexpandnonlinear1}
    &p_{\hat{i}}(t)|h_{{\hat{i}}i}|^2 + 2 \alpha |h_{{\hat{i}}i}| \sum_{n=1}^N p_{\hat{i}}(t)\epsilon_{i,n}(t) \mathbf{H}_{i,{\hat{i}},n} + \nonumber\\
    &\alpha^2 \bigg( \sum_{n=1}^N \underbrace{\Tilde{p}_{\hat{i},i,n}(t)}_\text{linear1} \mathbf{H}_{i,{\hat{i}},n}^2 + \sum_{x=1}^N \sum_{\substack{y=1\\ y \neq x}}^N \underbrace{\tilde{\Tilde{p}}_{\hat{i},i,xy}(t)}_\text{linear2} \mathbf{H}_{i,{\hat{i}},x}\mathbf{H}_{i,{\hat{i}},y} \bigg) \nonumber\\
    &\geq (2^{R_{th}}-1){\sigma^2}
\end{align}
\normalsize

where $\Tilde{p}_{\hat{i},i,n}(t)= p_{\hat{i}}(t)\epsilon_{i,n}(t)$ is linearized by the three inequalities \eqref{ineq1}-\eqref{ineq3} when replacing $\Tilde{p}_{\hat{i},z,n}(t)$ by $\Tilde{p}_{\hat{i},i,n}(t)$ and  $\xi_{z,n}(t)$ by $\epsilon_{i,n}(t)$. Similarly, $\tilde{\Tilde{p}}_{\hat{i},i,xy}(t)=p_{\hat{i}}(t)\epsilon_{xy,n}(t)$ is linearized by the three inequalities \eqref{ineq1}-\eqref{ineq3} when replacing ${\Tilde{p}}_{\hat{i},z,n}(t)$ by $\tilde{\Tilde{p}}_{\hat{i},i,xy}(t)$ and $\xi_{z,n}(t)$ by $\epsilon_{xy,n}(t)$.

Now the non-linear constraint \eqref{rate} can be represented as a linear constraint by considering \eqref{rateexpandnonlinear1}. Then the linear version of the optimization problem can be rewritten as:

\small
\begin{equation}\label{R maximization2a}
    \underset{\substack{\epsilon_{i,n}(t), \epsilon_{i,x(t)}, \epsilon_{i,y}(t), \tilde{\epsilon}_{i,xy}(t), \\ p_i(t) \geq 1, \Tilde{p}_{\hat{i},i,n}(t), \tilde{\Tilde{p}}_{\hat{i},i,xy}(t)}}{\text{minimize}} \quad \zeta E^\text{RIS}_C(t) + (1-\zeta) \sum\limits_{i=1}^2 E^U_i(t)
\end{equation}
subject to: \newline

     ~~~~~~~~~~~\eqref{peakP}, \eqref{assoc}, \eqref{energy1}, \eqref{energy2}, \eqref{lineartwobinary1}-\eqref{lineartwobinary3}, \eqref{rateexpandnonlinear1}
\begin{equation}\label{ineq1}
    {p}_i(t) \geq \Tilde{p}_{\hat{i},i,n}(t) \geq 0,
\end{equation}
\begin{equation}\label{ineq2}
    \Tilde{p}_{\hat{i},i,n}(t) \geq P^\text{max}\epsilon_{i,n}(t) - P^\text{max} + {p}_i(t)
\end{equation}
\begin{equation}\label{ineq3}
    \Tilde{p}_{\hat{i},i,n}(t) \leq P^\text{max}\epsilon_{i,n}(t)
\end{equation}
\begin{equation}\label{ineq1}
    {p}_i(t) \geq \tilde{\Tilde{p}}_{\hat{i},i,xy}(t) \geq 0,
\end{equation}
\begin{equation}\label{ineq2}
    \tilde{\Tilde{p}}_{\hat{i},i,xy}(t) \geq P^\text{max}\tilde{\epsilon}_{i,xy}(t) - P^\text{max} + {p}_i(t)
\end{equation}
\begin{equation}\label{ineq3}
    \tilde{\Tilde{p}}_{\hat{i},i,xy}(t) \leq P^\text{max}\tilde{\epsilon}_{i,xy}(t)
\end{equation}
\normalsize

\subsection{Bender Decomposition}
Although the optimization problem can be solved optimally based on the solution in Section (\ref{optimalsolution}), this solution is not efficient due to the presence of large number of binary variables. In this section, we propose to use Bender Decomposition (BD) to solve the optimization problem efficiently, in return, the solution is not guaranteed to be optimal. 
%rephrase
The BD decomposes the problem into two simpler problems, namely the master problem (MP) and the auxiliary problem (AUXP). The MP is a relaxed version of the original problem, containing only a
subset of the original variables, in our problem, it will contain the binary variable, and the associated constraints. Its solution yields a lower bound on the objective function. The AUXP is the original problem with the variables obtained in the MP fixed, i.e., $\epsilon_{i,n}(t)$. Its solution yields an upper bound on the objective function and is used to generate cuts for the MP. The MP and AUXP
are solved iteratively, until the upper and lower bounds are sufficiently close \cite{BD}.

Given that we have two decision variables, $p_i(t)$ and $\epsilon_{i,n}(t)$, we are going to fix $p_i(t)$ in the MP and solve for $\epsilon_{i,n}(t)$ then we are going to do the opposite in the auxiliary problem.

\subsubsection{Main Problem (MP)}

By fixing $p_i(t)$ to a given value $\bar{p}_i(t)$, the MP is given as follows:

\begin{equation}\label{R maximization222}
    \underset{\epsilon_{i,n}(t)}{\text{minimize}} \quad  E^\text{RIS}_C(t) 
\end{equation}
subject to:
  \begin{equation}\label{assoc22}
\sum\limits_{i=1}^ 2 \epsilon_{i,n}(t) \leq 1, \quad \forall n, \forall t,
 \end{equation}
  \begin{equation}\label{rate22}
R_i(t) \geq R_{th} \quad \forall t, \forall i=1,2.
 \end{equation}
  \begin{equation}\label{energy122}
E^\text{RIS}_C(t) \leq E_S(t-1), \forall t,
 \end{equation}
  \begin{equation}\label{energy222}
E_S(t-1)+E_H(t) \leq \bar{E}_S, \forall t,
 \end{equation}

The second term of the objective function is eliminated because it is now constant with respect to $p_i(t)$. Also, constraint (\ref{peakP}) is not considered in this MP for the same reason. Checking the linearity of this optimization problem, we will find that the objective function and all constraints are linear except constraint (\ref{rate22}) which can be written as follows:

\small
\begin{equation}
     \bigg({\bar{p}_{\hat{i}}(t)(|h_{{\hat{i}}i}|+\alpha \sum_{n=1}^N \epsilon_{i,n}(t) \mathbf{H}_{i,{\hat{i}},n})^2}\bigg) \geq (2^{R_{th}}-1){\sigma^2}
\end{equation}
\normalsize

\small
\begin{equation}
     (|h_{{\hat{i}}i}|+\alpha \sum_{n=1}^N \epsilon_{i,n}(t) \mathbf{H}_{i,{\hat{i}},n})^2 \geq \frac{(2^{R_{th}}-1){\sigma^2}}{\bar{p}_{\hat{i}}(t)}
\end{equation}
\normalsize

Taking square root both sides yields to:

\small
\begin{equation}
     (|h_{{\hat{i}}i}|+\alpha \sum_{n=1}^N \epsilon_{i,n}(t) \mathbf{H}_{i,{\hat{i}},n}) \geq
     \sqrt{\frac{(2^{R_{th}}-1){\sigma^2}}{\bar{p}_{\hat{i}}(t)}}
\end{equation}
\normalsize

The resulting equation above is linear in $\epsilon_{i,n}(t)$ and, hence, the MP is now linear and can be solved as an Integer Linear Programming (ILP).

\subsubsection{Auxiliary Problem (AUXP)}
Similar to the MP, by fixing $\epsilon_{i,n}(t)$ to a given value $\bar{\epsilon}_{i,n}(t)$, the AUXP can be given as:

\begin{equation}\label{R maximization233}
    \underset{p_i(t) \geq 1}{\text{minimize}} \quad \sum\limits_{i=1}^2 E^U_i(t)
\end{equation}
subject to:
 \begin{equation}\label{peakP33}
 p_i(t) \leq P^\text{max}, \quad \forall t, \forall i=1,2.
 \end{equation}
  \begin{equation}\label{rate33}
R_i(t) \geq R_{th} \quad \forall t, \forall i=1,2.
 \end{equation}
  \begin{equation}\label{energy133}
E^\text{RIS}_C(t) \leq E_S(t-1), \forall t,
 \end{equation}
  \begin{equation}\label{energy233}
E_S(t-1)+E_H(t) \leq \bar{E}_S, \forall t,
 \end{equation}

The first term of the objective function is eliminated because it is now constant with respect to $\epsilon_{i,n}(t)$. Also, constraint (\ref{assoc}) is not considered in this auxiliary problem for the same reason. Checking the linearity of this optimization problem, we will find that the objective function and all constraints are linear except constraint (\ref{rate33}) which can be written as follows:

\small
\begin{equation}\label{optimalP}
     p_{\hat{i}}(t) \geq \frac{(2^{R_{th}}-1){\sigma^2}}{(|h_{{\hat{i}}i}|+\alpha \sum_{n=1}^N \bar{\epsilon}_{i,n}(t) \mathbf{H}_{i,{\hat{i}},n})^2}
\end{equation}
\normalsize

Note that this constraint is now linear in $p_{\hat{i}}(t)$. Given that the auxiliary optimization problem's objective is to minimize the transmit power of both users, the optimal value of $p_{\hat{i}}(t)$, denoted as $p^*_{\hat{i}}(t)$ can be obtained by changing the inequality of constraint \eqref{optimalP} to equality.

\section{Simulation Results}

We consider an RIS serving a couple of users, $u_1$ and $u_2$, located at coordinates (1, 1) and (1000, 1), respectively. We vary the $x$ coordinate of the RIS in order to study the optimal locations under different scenarios. The $x$ coordinates of the RIS ranges from 1 to 1000, and the $y$ coordinate is 150. Regarding energy harvesting, we assume that $\theta(t)$ follows a truncated normal distribution with mean 2 W and variance 0.25 in the interval [0, 2.4]. In our simulation, we calculate the average solution over 35 different scenarios. The rest of the simulation parameters are shown in Table \ref{table1}.

\begin{table}[h!]
\caption{Simulation Parameters}
\centering
\addtolength{\tabcolsep}{1pt}\begin{tabular}{|c |c ||c |c|}
\hline
Parameter    & Value & Parameter & Value \\ [0.5ex]
  \hline \hline
$P^\text{max}$   &   1 W    &    $P_e$    & 5 mW   \\
\hline
N          &   50  &      $T_s$       & 1  \\
\hline
$\sigma^2$     &   3.9811e-11  &  $\eta$   &  0.9 \\
\hline
Path loss constant     &   2  &       $n$  & 50 \\
\hline
$\alpha$      &    1     & $R_{th}$            &  0.1, 5, 7, 7.5  \\
\hline
\end{tabular}
\label{table1}
\end{table}

\begin{figure}[h!]
  \centerline{\includegraphics[width=3in]{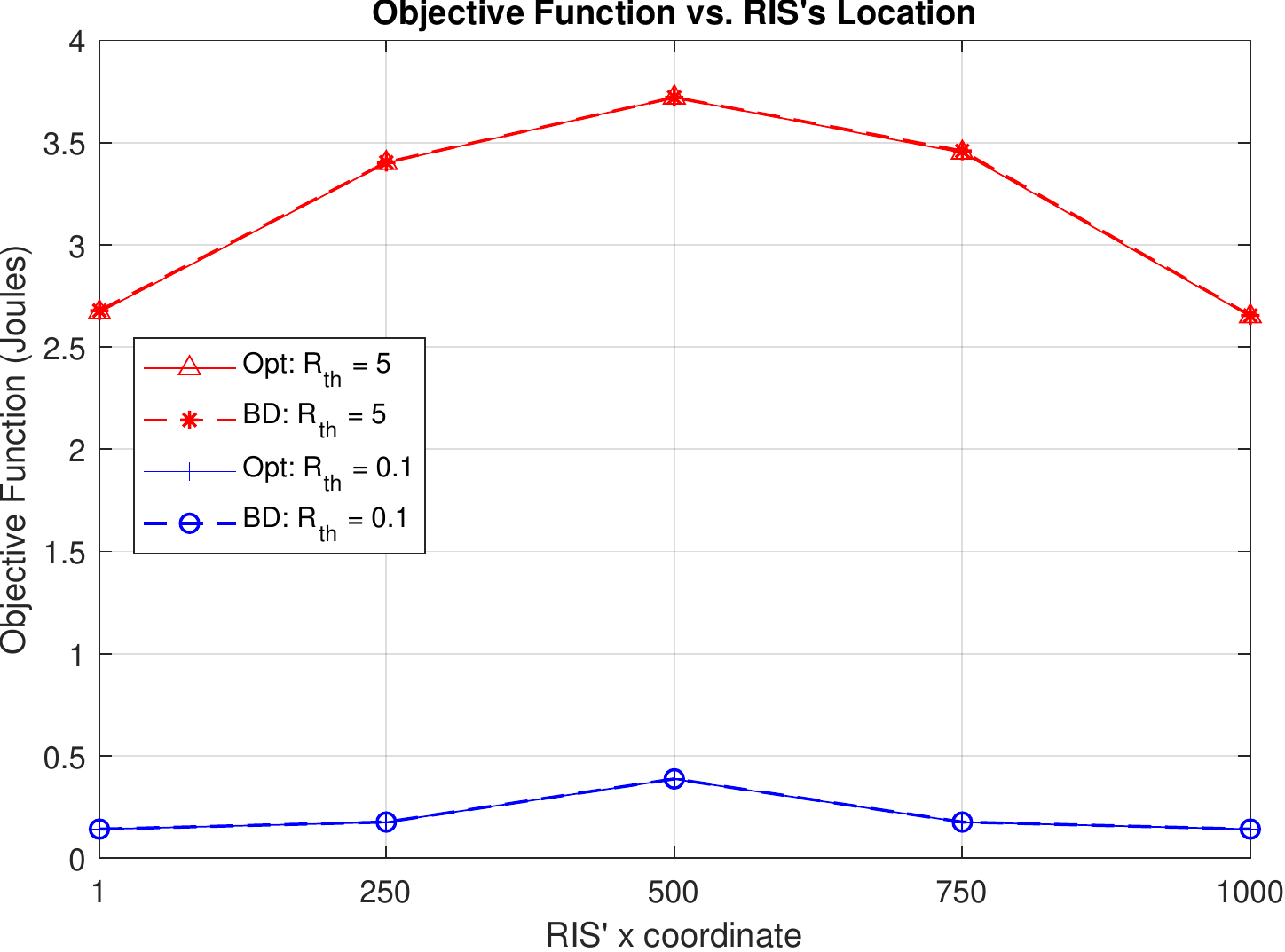}}
   \caption{Optimal vs. BD Solutions with different $R_{th}$ values and RIS's locations.}\label{fig2}
\end{figure}
In Fig. \ref{fig2}, we compare the optimal and BD solutions with different $R_{th}$ values and RIS's locations. The objective is to minimize energy consumed by the users. It is shown that Bender Decomposition achieves a close solution to the optimal solution. On the other hand, we can see that the highest total energy consumption happens when the RIS is located in the middle. This is because both users need to reach the RIS that is located relatively far away from them.

\begin{figure}[h!]
  \centerline{\includegraphics[width=3in]{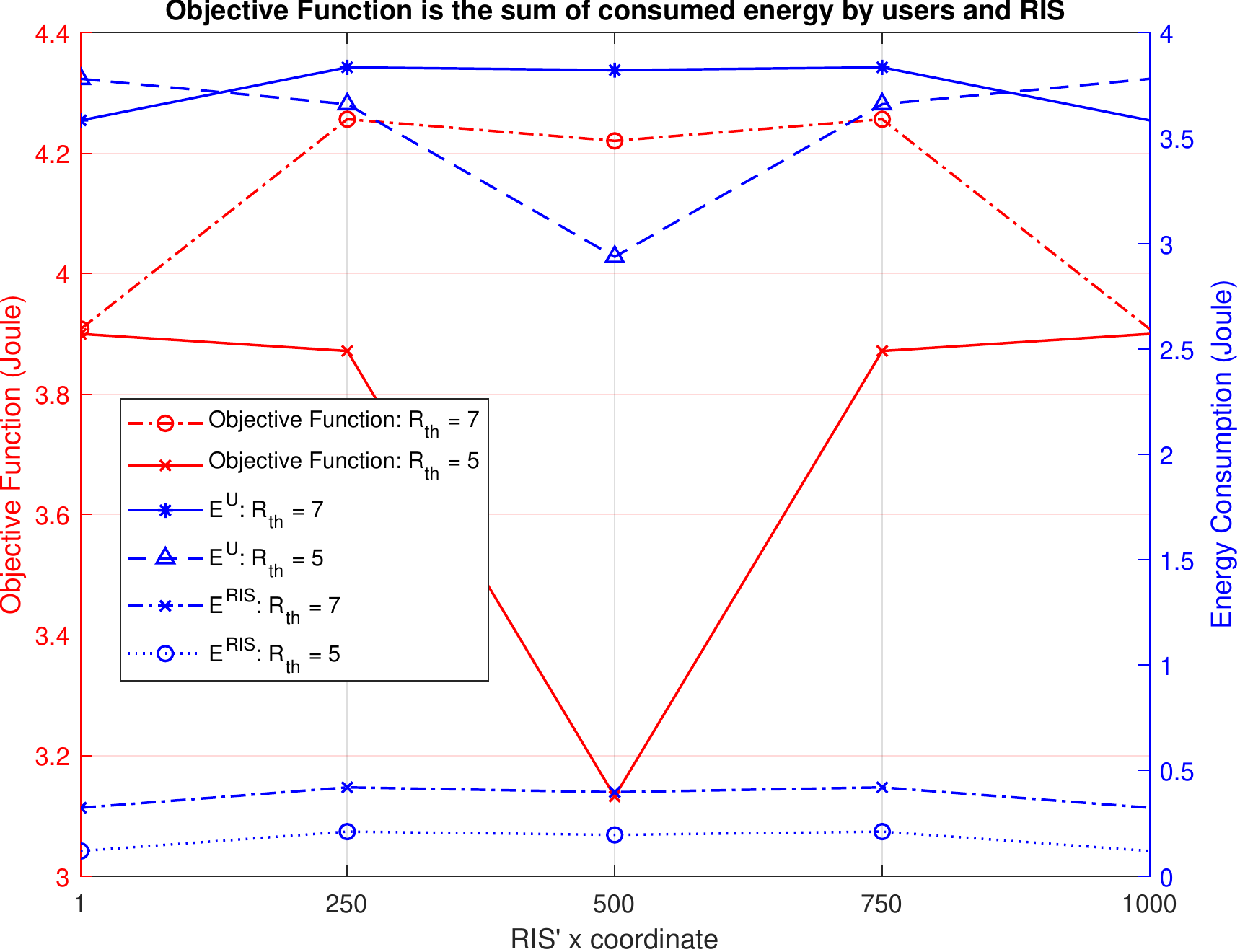}}
   \caption{Sum of energy consumed by the RIS and users vs. different $R_{th}$ values and RIS's locations.}\label{fig3}
\end{figure}
\begin{figure}[h!]
  \centerline{\includegraphics[width=3in]{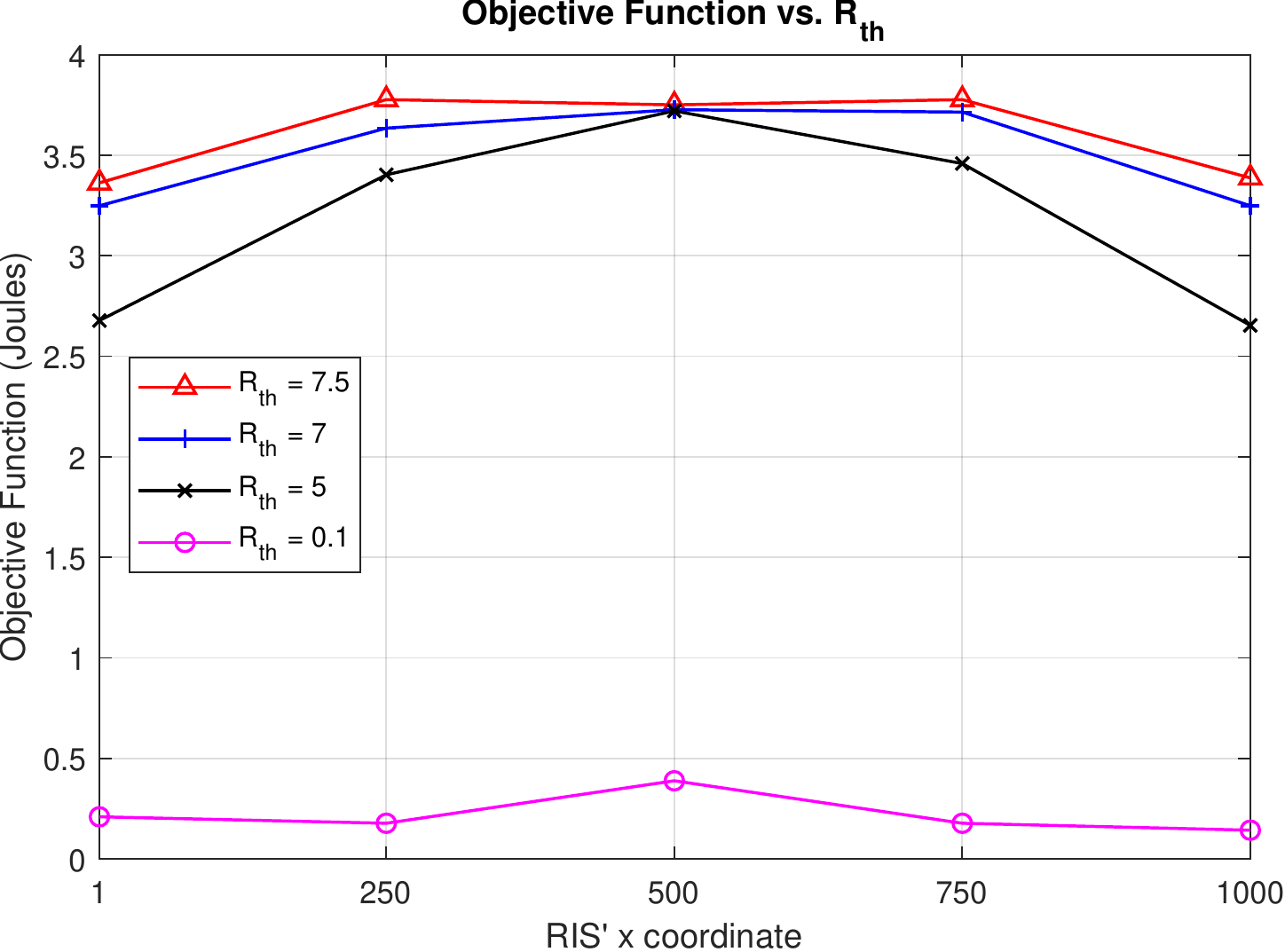}}
   \caption{Energy consumed by users vs. different $R_{th}$ values and RIS's locations.}\label{fig4}
\end{figure}
\begin{figure}[h!]
  \centerline{\includegraphics[width=3in]{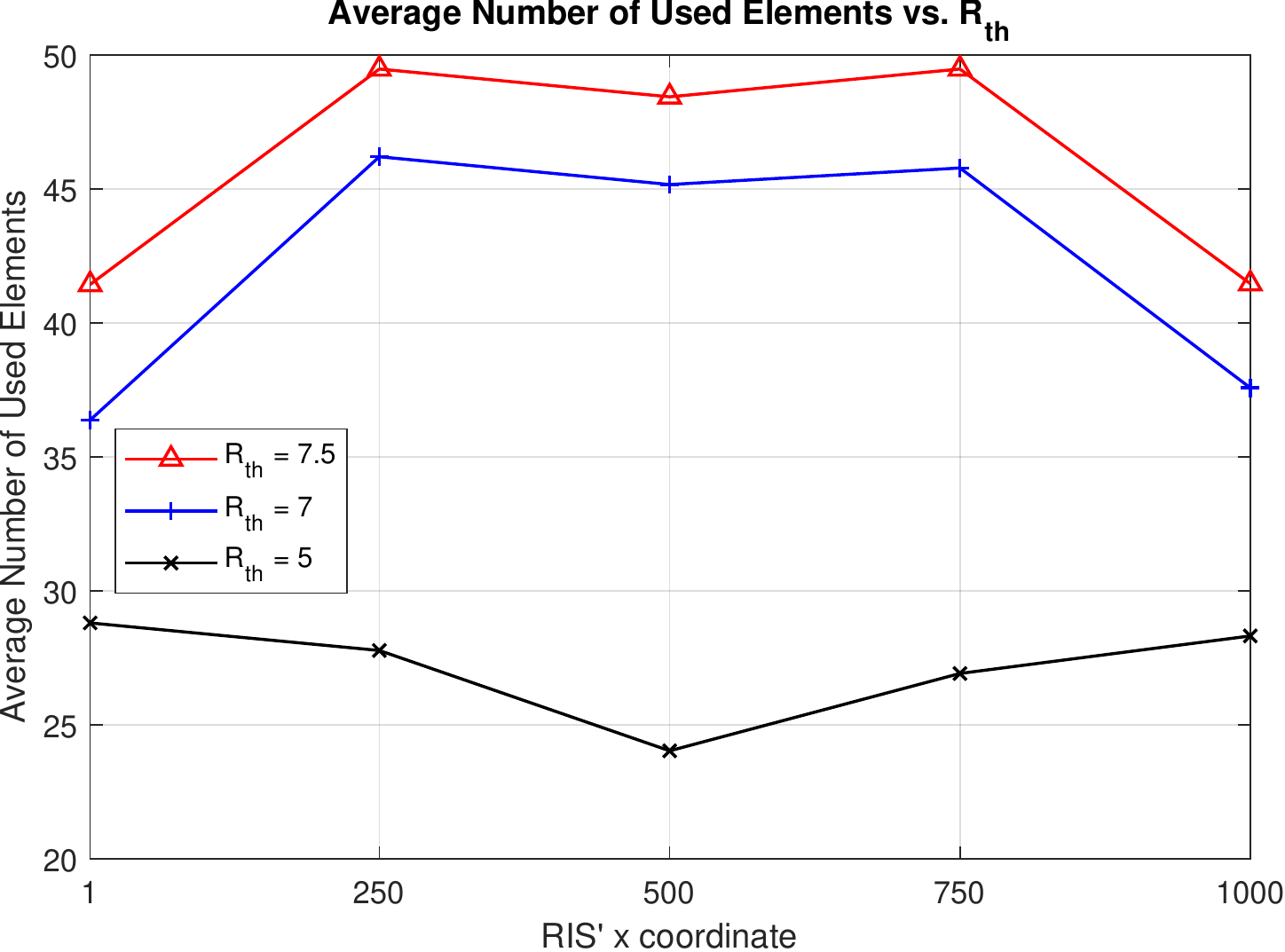}}
   \caption{Average number of used RIS's elements vs. different $R_{th}$ values and RIS's locations.}\label{fig5}
\end{figure}

Fig. \ref{fig3} illustrates energy consumption, under different $R_{th}$ values and RIS locations, when the objective is to minimize the total energy consumed by the users and the RIS. Fig. \ref{fig5} indicates that the total energy consumption is dominated by the users' energy consumption. The users need to transmit with high powers to reach the RIS since the RIS is either far away from one user or in the middle but relatively not close to both of them. Therefore, we investigate energy consumption when the objective is minimizing the users' energy consumption, as shown in Fig. \ref{fig4} and Fig. \ref{fig5}.

Fig. \ref{fig4} presents the effect of varying $R_{th}$ to the total energy consumption of the users. It is shown that increasing the value of $R_{th}$ impacts energy consumption of the users. A higher value for $R_{th}$ results in a higher energy consumption. The reason is that increasing the threshold $R_{th}$ increase the data rate function, as described in constrain (\ref{rate}). Accordingly, this can lead to increasing the transmission powers of the users, i.e., energy consumption.

Fig. \ref{fig5} shows the average number of used RIS reflecting elements under different $R_{th}$ thresholds and RIS locations. A higher value for $R_{th}$ forces either the users to increase their transmission powers and/or the RIS to exploit more reflecting elements in order to satisfy the data rate requirement. This is because the data rate is a function of transmission power and used reflecting elements. Consequently, RIS tends to utilize more reflecting elements as the data rate requirement of the users increases.

When the value of $R_{th}$ is 5, the RIS tends to utilize more reflecting elements when it is located close to the location of either users, where the first and second users are located at coordinates (1, 1) and (1000, 1), respectively. In this case, one user is far away from the RIS, and hence, utilizing more reflecting elements to support this user is needed. As we move the RIS toward the middle location between the two users (when the $x$ coordinate of the RIS is 500), the users are already required to transmit with a higher power to reach the RIS, and the RIS does not have to utilize more reflecting elements where the required data rate demand can be supported. Interestingly, the RIS is obligated to utilize more reflecting elements as it move toward the middle point between the users when the value of $R_{th}$ is 7 or 7.5. The reason is that the users need to transmit with high power to support a higher data rate, which could not be satisfied without utilizing more reflecting elements by the RIS.

\begin{figure}[h!]
  \centerline{\includegraphics[width=3in]{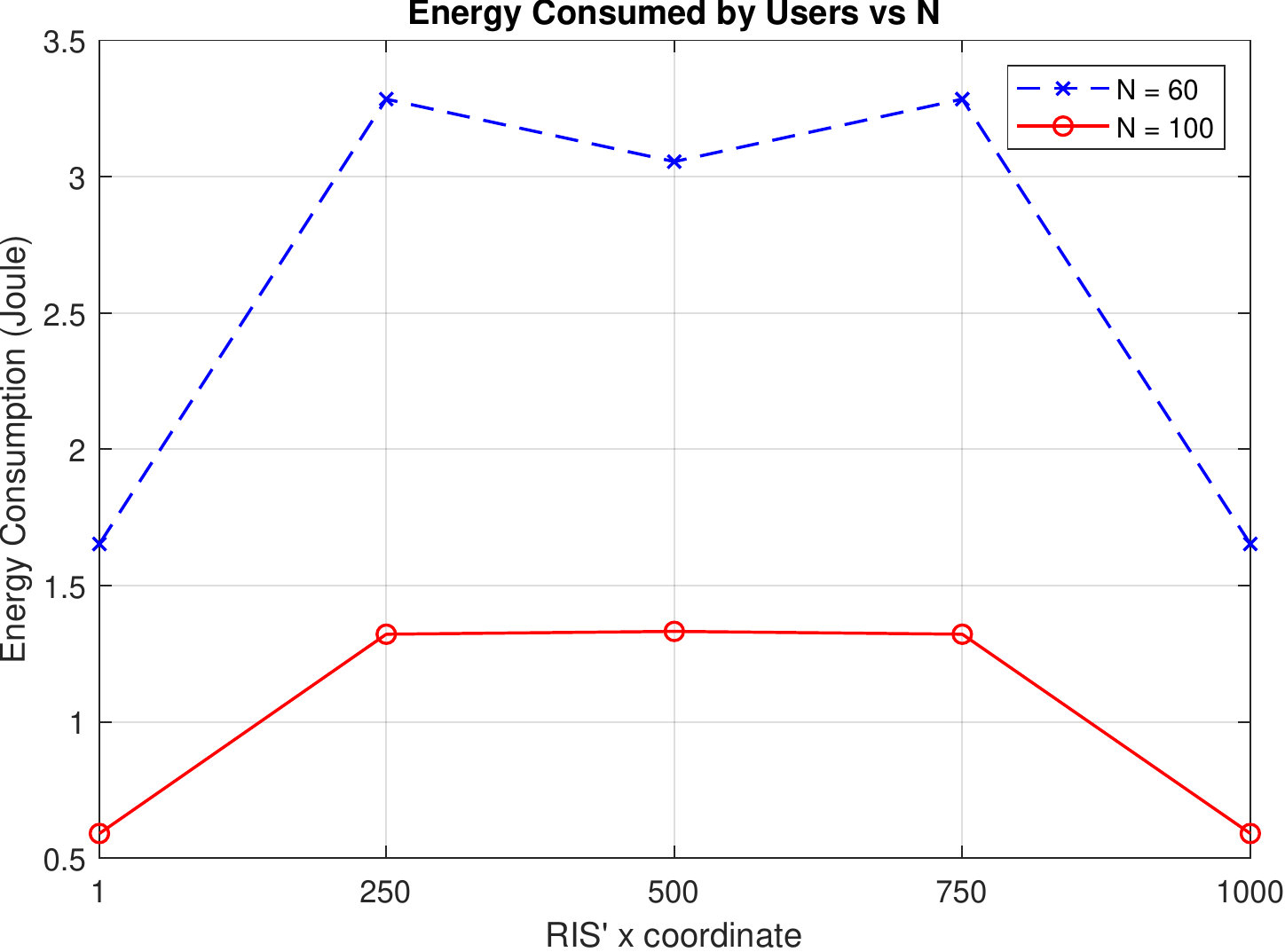}}
   \caption{Energy consumed by users vs. $N$ and RIS's locations. }\label{fig6}
\end{figure}
Fig. \ref{fig5} presents the effect of the total number of reflecting elements on the total energy consumed by users. Total energy consumed by the user when $N = 100$ is less than the energy consumed by them when $N=50$, where $R_{th}$ = 1 and $P^{max}$ = 2 W. It is shown that more reflecting elements available to the users can contribute in reducing the total energy consumption of the users. For a given $R_{th}$, the user can achieve the data rate goal while transmitting with a lower power level when the number of the available reflecting element is higher, as described by equation (\ref{Rate_threshold}).

\section{Conclusions}
This paper investigates the use of an RIS panel to improve bi-directional communications between two user, and these can be end users or a basestation and end user. 
We considered the case in which the RIS panel is powered using a solar panel that harvests energy, and the harvested energy powers the reflecting elements as well as the panel's microcontroller(s).
An optimization problem was formulated to decide the transmit power of each of the two users, and the number of elements that will be used to reflect the signal of any two communicating pairs in the system while guaranteeing a minimum transmission rate. 
The optimization problem assumes that the RIS panel is located at a given location, and is a mixed-integer non-linear problem, and its objective is the minimization of the energy consumed by the end users.
The non-linear constraints are minimized, and a heuristic approach using the Bender decomposition is used to find a near optimal solution.
Numerical results showed that the proposed model is capable of delivering the minimum rate of each user even if line-of-sight communication is not achievable.
It also showed that the energy consumption depends on the location of the RIS panel.

%\section{Mohamed Backhaul Outage Compensation}
%\input{Mohamed Backhaul Outage Compensation}

%\appendices
%\section{Concavity Proof}
\bibliographystyle{IEEETran}
\bibliography{TWC_IRS.bib}
\end{document}